\documentclass[journal=jacsat,manuscript=article]{achemso}

\usepackage{chemformula}
\usepackage[T1]{fontenc}
\usepackage{comment}
\usepackage{algorithm}
\usepackage{rotating}
\usepackage[noend]{algpseudocode}
\usepackage{tikz}
\usetikzlibrary{shapes, arrows}
\usepackage{caption}
\usepackage{subcaption}
\usepackage[font=small]{caption}
\usepackage{amsfonts}
\usepackage{amsmath}
\usepackage{mathtools}
\usepackage{ulem}
\usepackage{bigints}
\usepackage{float}

\usepackage{hyperref}
\usepackage{graphicx}
\usepackage{caption}
\usepackage{braket}

\usepackage{soul} 

\usepackage{gensymb}

\usepackage{tcolorbox}
\definecolor{my-yellow}{HTML}{FFFDB8}
\definecolor{dark-yellow}{HTML}{F2BE22}
\definecolor{my-blue}{HTML}{D4FAFC}
\definecolor{dark-blue}{HTML}{22A699}

\graphicspath{ {./images/} }

\title{An improved guess for the variational calculation of charge-transfer excitations in large systems}

\author{Nicola Bogo}

\affiliation[TUM]
{Department of Chemistry and Catalysis Research Center, TUM School of Natural Sciences, Technische Universität München, Lichtenbergstr. 4, 85748 Garching, Germany}

\author{Zeyi Zhang}

\affiliation[UC Berkeley]
{Pitzer Center for Theoretical Chemistry, Department of Chemistry, University of California, Berkeley, California 94720, United States.}

\alsoaffiliation[LBL]
{Chemical Sciences Division, Lawrence Berkeley National Laboratory, Berkeley CA 94720, USA.}

\author{Martin Head-Gordon}

\affiliation[UC Berkeley]
{Pitzer Center for Theoretical Chemistry, Department of Chemistry, University of California, Berkeley, California 94720, United States.}

\alsoaffiliation[LBL]
{Chemical Sciences Division, Lawrence Berkeley National Laboratory, Berkeley CA 94720, USA.}

\author{Christopher J. Stein}
\email{christopher.stein@tum.de}

\affiliation[TUM]
{Department of Chemistry and Catalysis Research Center, TUM School of Natural Sciences, Technische Universität München, Lichtenbergstr. 4, 85748 Garching, Germany.}

\alsoaffiliation[AMC]
{Atomistic Modeling Center, Munich Data Science Institute, Technical University of Munich, Garching 85748, Germany.}

\SectionNumbersOn

\begin{document}

\maketitle

\begin{abstract}
    Charge-transfer excited states are highly relevant for applications in molecular electronics. However, the accurate calculation of these states in large systems is challenging since wave function methods are prohibitively expensive, time-dependent density functional theory with typical functionals is not precise, and the complicated topology of the electronic hypersurface makes the variational convergence to the targeted excited states a difficult task. We address the latter aspect by providing suitable initial guesses which we obtain by two separate constrained algorithms. Combined with subsequent squared-gradient minimization schemes, we demonstrate that OO-DFT calculations can reliably converge to the charge-transfer states of interest even for large molecular systems. We test this approach on two chemically very different supramolecular structures and also analyze the performance of two recently proposed methods for the tuning of the range-separation parameter in time-dependent DFT with range-separated hybrid functionals. Our results demonstrate that with the methods presented here, reliable convergence of charge-transfer excited states can be achieved with variational excited-state DFT methods, while time-dependent DFT calculations with an adequate tuning procedure for the range-separation parameter can provide a computationally efficient initial estimate of the corresponding energies.
\end{abstract}

\section{Introduction}
Over the last decade, the resolution of photo-lithographic techniques for silicon semiconductor fabrication has approached the size of molecules, and the physical limits of optical technology have halted further miniaturization. A bottom-up approach where molecular components are assembled into molecular devices, like biological photosynthetic systems, would allow further miniaturization in chip design, paving the way for a disruptive innovation in semiconductor technology: carbon-based semiconductors. However, the cost of organic building blocks equipped with extended $\pi$ electrons systems, which are delocalized and prone to participate in conduction, the development costs of novel synthesis and characterization techniques, and ultimately, the impact of trial and error on semiconductor design amount to a significant barrier in the development of these materials. \\
A rational strategy for the development of carbon-based semiconductor technology implies the study of charge transfer (CT) in biological systems, its reproduction in bio-inspired energy materials, and their technological application. The fundamental understanding of charge-transfer processes enabled by theoretical work is fundamental, forging guiding principles for molecular design and testing them with computer simulations. Given this premise, excited-state electronic structure methods play a crucial role for the study and discovery of charge transport in molecular systems. Since their introduction in the 1990s, tools based on density functional theory (DFT) have achieved predictive power in the simulation of ground-state (GS) chemistry and are routinely used to develop new chemicals. Likewise, excited-state electronic-structure methods based on DFT, like time-dependent density functional theory\cite{runge1984density} (TD-DFT), gained popularity due to the straightforward simulation of light absorption features in small molecules. In a TD-DFT calculation, an approximation of the response function of the system under investigation is computed. The success of TD-DFT resides in the application of efficient approximations that lower its computational demand at the cost of introducing small errors. With a focus on economic simulations, the most popular TD-DFT methods expand the response function in a Taylor series and truncate it to the linear term, applying the Tamm--Dancoff approximation\cite{hirata1999time} (TDA) to recast the resulting electronic structure problem to the well-known configuration-interaction singles\cite{bene1971self, foresman1992toward} (CIS) method. The resulting method, which we name TDA from here on, is plagued by systematic errors when general-purpose exchange-correlation functionals (XCFs) are applied to the calculation of CT excitations. The proposed solutions consist of the application of more complicated XCF approximations\cite{yanai2004new, grimme2006semiempirical, mester2021simple},
increasing the computational cost of the electronic structure and often requiring system-specific parameter tuning\cite{baer2010tuned, prokopiou2022optimal}
for accurate results. \\
In our previous publication\cite{D4CP01866D}, we demonstrated how orbital-optimized DFT (OO-DFT) methods\cite{hait2020excited, hait2021orbital} constitute the most accurate low-scaling electronic structure approximation for the calculation of inter-molecular CT excitations using a general-purpose XCF. 
OO-DFT methods compute variationally the ground state and an electronically-excited configuration where two separate variational calculations are performed on the ground and excited state. 
A rigorous theoretical foundation has recently been proven for $\Delta$SCF,\cite{yang2024foundation} whereas the application of variational DFT methods to excited electron configurations was previously motivated by the promising results showcased by its computational implementations\cite{hait2020highly, selenius2024orbital, barca2018simple}. 
OO-DFT methods leverage numerical optimization to compute the stationary points of the electronic energy hypersurface, closely resembling ground-state DFT methods that minimize the electronic energy. 
Conversely, excited-state energies belong to saddle points of the electronic hypersurface, making numerical convergence more challenging. Indeed, we relied on outlier detection rules to exclude data points flawed by poor convergence in our benchmark study, where we subsequently tested the accuracy of OO-DFT methods. 
Schmerwitz \textit{et al.}\cite{schmerwitz2025freeze} analyzed their freeze-and-release optimization algorithm on the donor-acceptor scans from our $R_\mathrm{DA}$-dataset, demonstrating the importance of a well-behaved guess for successful OO-DFT calculations. 
In this contribution, we first compare guess-refinement strategies and OO-DFT algorithms implemented in the Q-Chem software package\cite{epifanovsky2021software} on the intermolecular charge-transfer (ICT) excitations in the tetrafluoroethylene-ethylene dimer system from the $R_\mathrm{DA}$-dataset, that we consider a good benchmark for convergence of OO-DFT methods. We implement and analyze a constrained optimization algorithm for OO-DFT guess refinement method using the efficient geometric direct minimization (GDM) algorithm\cite{van2002geometric}, akin to the approach introduced by Schmerwitz \textit{et al}.\cite{van2002geometric} 
Moreover, we investigate an alternative guess for ICT excitations obtained from the absolutely-localized molecular orbitals\cite{stoll1980use, gianinetti1996modification} (ALMO) method and compare the maximum overlap method to an initial guess\cite{barca2018simple} (IMOM) to the squared-gradient minimization\cite{hait2020excited, hait2021orbital} (SGM) algorithm on these refined guesses.\\
TDA is certainly of fundamental importance for the calculation of light absorption properties, and the OO-DFT methods we apply in this work also require an initial TDA calculation. Nevertheless, results depend significantly on the chosen XCF approximation and should therefore be validated for a given compound class by comparing them to reference methods. This is not always possible because of the computational scaling of wavefunction methods. Due to this concern, we applied OO-DFT methods to CT excitations in prototypical photo-induced charge transfer (PCT) systems with applications in the field of organic semiconductors. \\
In section \ref{sec:methods}, we review the current state-of-the-art methods for variational calculations of electronically excited states. Our implementation of an efficient partitioning of the MO matrix $\mathbf{C}$ is discussed in section \ref{sec:frzopt}. Convergence and accuracy of the SGM and IMOM methods using the ground-state MO guess, ALMO guess, and refined MO guess through constrained optimization (FR guess) are tested on the tetrafluoroethylene-ethylene dimer in section \ref{sec:tfetet}. Finally, the guess refinement algorithm is combined with the SGM algorithm for the calculation of CT excitations in large systems. Low-lying PCT excitations involving a phenothiazine dye as electron donor and anthraquinone dye as acceptor in a Pd coordination cage\cite{frank2016light} are computed in section \ref{sec:clevercage}. CT excitations in dye-semiconductor complexes\cite{gemeri2022electronic} are computed in section \ref{sec:dye-TiO2}. The results obtained with SGM are compared to the TDA prediction obtained with recent reparametrizations of the LRC-$\omega$PBE\cite{rohrdanz2008simultaneous} XCF for the TDA method and experimental measurements.

\section{Methods}\label{sec:methods}

A rigorous introduction to TD-DFT is beyond the scope of our study, so we point the reader to an exhaustive introduction\cite{marques2012fundamentals}, and focus instead on the variational methods we employed. 
In the following, Greek indices denote the Gaussian AO basis functions in our notation, and Latin characters denote the MOs. 
We use indices $j, k$ for occupied MOs, $a, b$ for virtual MOs, and $m, n$ for general MOs, independent of occupation.
The superscript $(i)$ indicates the iteration number in variational electronic-structure calculations.
For ground-state calculations, the convergence of the electronic structure is accelerated with two alternative approaches: direct optimization (DO) and extrapolation methods. 
DO methods apply a unitary transformation $\mathbf{U}$ to the MO matrix $\mathbf{C}$ at every step
\begin{align}
\mathbf{C}^{(i+1)} = \mathbf{C}^{(i)} \mathbf{U}\, .
\end{align}
The unitary transformation is parametrized by taking the exponent of an antisymmetric matrix $\Delta$
\begin{align}
\mathbf{U} = e^\Delta.
\end{align}
This way, the optimization is performed directly on the elements of $\Delta$, instead of applying small variations on the $\{c_{i\mu}\}$ elements while constraining the columns of $\mathbf{C}$ to stay orthonormal. 
The energy minimization involves the evaluation of the gradient of the electronic energy with respect to the upper triangular elements of $\Delta$. Suitable minimization algorithms are \textit{e.g.}  Broyden--Fletcher--Goldfarb--Shanno (BFGS) which is used in the implementation of the geometric direct minimization\cite{van2002geometric} (GDM) algorithm in Q-Chem. 
The elements of $\Delta$ are the orbital rotations denoted by the letter $\theta$. $\theta_{mn}^{(i)}$ mixes MOs $m$ and $n$ from the $i$-th iteration to construct $\mathbf{C}$ in iteration $i+1$, and only rotations mixing occupied and virtual MOs $\theta_{ja}^{(i)}$ (off-diagonal blocks in $\Delta$) affect the total electronic energy.\\
Conversely, extrapolation methods, such as the direct inversion of the iterative subspace (DIIS) method from Pulay\cite{pulay1980convergence, pulay1982improved}, leverage the mathematical properties of the Fock $\mathbf{F}^{(i)}$ and density $\mathbf{P}^{(i)}$ matrices at convergence to construct an error vector. 
The commutator between density and Fock matrix vanishes at convergence, such that 
\begin{align}
\mathbf{e}^{(i)} = \mathbf{S} \mathbf{P}^{(i)} \mathbf{F}^{(i)} - \mathbf{F}^{(i)} \mathbf{P}^{(i)} \mathbf{S}
\end{align}
is a suitable error vector where
the overlap matrix is denoted as $\mathbf{S}$. 
In DIIS, the density matrix is extrapolated with information from previous iterations. As a consequence, it is not guaranteed that the closest local minimum is found for a given initial guess since the extrapolation can lead to tunneling through the local basin.
In case of ground-state calculations, the combined DIIS-GDM algorithm accelerates convergence by using DIIS in the first iterations until the DIIS error is below a given threshold and refining the DIIS guess with GDM to tight convergence.
Excited states, however, are saddle points instead of minima. An adaptation of the algorithms to a variational optimization of excited states is, therefore, not straightforward. \\
To mitigate the complexity of direct saddle-point optimization of the electronic hypersurface, a constraint is applied to the density in constrained DFT\cite{kaduk2012constrained} (CDFT) methods. 
Constraints can impose site occupation\cite{goldey2017charge} conditions or can be applied directly to the density difference between ground and excited state\cite{kussmann2024constraint}. 
These methods hold great potential when enough information is available about the target excited state (\textit{e.g.} information about site occupancy), but in this work, we use constraints to compute an improved guess density. 
In our calculations, the electronic energy is hence computed by optimization of all degrees of freedom of the electronic energy without applying constraints. \\
To compute specific excited states, Gilbert \textit{et al.} proposed an adaptation of direct optimization or extrapolation algorithms\cite{gilbert2008self}, where the Fock matrix is constructed from the orbitals that overlap maximally with a reference configuration rather than following the Aufbau principle. 
In this maximum overlap method (MOM), the overlap of the occupied MO set belonging to the $i$-th iteration to a reference MO set $\mathbf{C}^{\mathrm{ref}}$
\begin{align}
\mathbf{O}=(\mathbf{C}^{\mathrm{ref}})^{\dagger}\mathbf{SC}^{(i)}
\end{align}
is computed at every iteration, where the element $O_{jk}$ quantifies the overlap between the $j$-th MO in the reference set and the $k$-th MO in the MO set belonging to the $i$-th iteration. 
The projection of the $k$-th MO onto the reference is  
\begin{align}
p_{k}=\left(\sum_{j}^{n}{O}_{j k}\right)^\frac{1}{2}
\end{align}
and the MOs are occupied starting from the highest occupation, to ensure the electronic structure is computed for the desired electron configuration.
As reference MOs, either the initial orbitals from a ground-state optimization can be used in the "maximum overlap to an initial guess" (IMOM) method,\cite{barca2018simple} or they can be updated for each iteration step (MOM).
The IMOM method is generally very fast and can swiftly converge to the targeted excited state if provided with a suitable MO guess. 
However, when the provided guess is far from the correct targeted state, convergence is challenging and undesired stationary points might be reached.
 The special case of CT excitations was investigated recently by Schmerwitz \textit{et al.},\cite{schmerwitz2025freeze} who showcased how the MOM method frequently fails to converge to the targeted stationary points of the energy hypersurface when coupled to a direct-optimization algorithm.
 The lack of convergence is the result of strong polarization in the excited-state density: over the iterations, the hole and particle MOs are mixed and the optimization algorithm cannot exit these strong mixing regions. \\
Hait and Head-Gordon proposed an alternative approach exploiting DO algorithms\cite{hait2020excited, hait2021orbital}, taking the square of the energy derivative with respect to occupied-virtual orbital rotations
\begin{align}\Delta=\left|\nabla_{\vec{\theta}}E\right|^{2}=\sum_{j a}\left|\frac{\partial E}{\partial\theta_{j a}}\right|^{2}
\end{align}
and feed this manipulated objective function to the DO algorithm. The resulting algorithm, named squared gradient minimization (SGM), which is usually coupled to standard line-search schemes, can converge on the saddle points of the electronic energy hypersurface.
Still, the manipulated objective function is characterized by undesired minima and cusps\cite{burton2022energy}, so it is again sensitive to the quality of the provided guess.
Moreover, in the SGM algorithm, the gradient is reduced quadratically when taking its square, meaning the condition number for the convergence must also be squared, requiring very tight convergence.
Despite these aspects, the SGM algorithm has been proven to be very successful \textit{e.g.} for the simulation of core-electrons spectroscopies\cite{hait2020highly} because ground or ionized states provide good guess functions for the orbital optimization, meaning the SGM algorithm starts in the appropriate quadratic well and can converge flawlessly.\\
Finally, Selenius \textit{et al.} proposed an efficient diagonal preconditioner for DO algorithms\cite{selenius2024orbital}
\begin{align}
{\frac{\partial^{2}E}{\partial\kappa_{m n}^{2}}}\approx-2\left(\epsilon_{m}-\epsilon_{n}\right)\left(f_{m}-f_{n}\right)\, ,
\end{align}
where $\epsilon_{m/n}$ are the orbital energies and $f_{m/n}$ are the occupation numbers of orbitals $m$ and $n$, respectively. This definition inverts the gradient along all the rotations mixing an occupied orbital with a lower-lying unoccupied orbital, ensuring the optimization is started by going ``uphill'' along the correct rotations that lead to the desired saddle point on the electronic hypersurface. 
This idea had already been employed for the construction of appropriate objective functions through manipulation of the gradient, employing efficient approximations of the inverse Hessian used in direct optimization algorithms.
This approach can be understood as an adaptation of the popular eigenvector-following algorithm for saddle-point search that is used for transition-state (TS) calculations to the electronic problem. 
A principal difference between the two applications is that only first-order saddle points are of interest for a TS search, whereas also higher-order saddle points are of interest for excited-state electronic structure.
This poses additional challenges for the variational calculations of electronic excitations. 
Like for TS calculations, eigenvector-following algorithms are sensitive to the quality of the initial guess.
Furthermore, molecular orbitals can change their energetic order over the DO iterations, leading to undesired consequences:
Firstly, energy terms belonging to the excited orbital directly enter the minimized electronic energy, whereas virtual orbitals don't, and this can guide the optimization algorithm into the aforementioned strong mixing regions with occupied-virtual rotations of 45\textdegree. Secondly, when the energetic order of occupied and virtual MOs is swapped, the target saddle point order is changed, and the objective function must be altered accordingly to avoid convergence to the wrong stationary point. \\
In the end, $\Delta$SCF methods are based on local optimization algorithms, and the underlying problem with variational calculations of excited-state energies resides in the complexity of the energy hypersurface.
Rapid convergence to the desired stationary point hence requires a suitable initial guess.
Therefore, guess-refinement algorithms are desirable to reduce the number of iterations and ensure convergence to the correct electronic configuration in OO-DFT calculations. 
Schmerwitz \textit{et al.} made use of constrained optimization in their variational excited-state methods\cite{levi2020variational, schmerwitz2023calculations, schmerwitz2025freeze} implemented in the GPAW software package\cite{mortensen2024gpaw} with broad success and proved the reliability of their approach by recomputing the outlier data points from our $R_\mathrm{DA}$-dataset. 
This motivated us to implement our variant of a constrained optimization algorithm and test the performance of the IMOM and SGM algorithms on the refined guess. 
Details about the constrained optimization algorithm are discussed in section \ref{sec:frzopt}. \\
An alternative method for the calculation of inter-molecular excitations is the absolutely-localized molecular orbitals (ALMO) method. 
ALMO enables the variational calculation of molecular fragments by partitioning the Gaussian basis $\{\varphi_\mu\}$ spanning the rows of the $\mathbf{C}$ matrix into localized subsets. In the simple case of two fragments, named A and B, the Gaussian basis is partitioned into the localized bases $\{\varphi_\nu\}^\mathrm{A}$ and $\{\varphi_\kappa\}^\mathrm{B}$ belonging to fragments A and B respectively. 
The MO matrix $\mathbf{C}$ is therefore enforced to maintain the block structure
\begin{align}
\mathbf{C}=\left[{\begin{array}{l l}{\mathbf{C}_\mathrm{A}}&{0}\\ {0}&{\mathbf{C}_\mathrm{B}}\end{array}}\right]
\end{align}
throughout the calculation, where $\mathbf{C}_\mathrm{A}$ and $\mathbf{C}_\mathrm{B}$ are the MO matrices of fragments A and B, respectively. The ALMO method is based on the equations of the locally-projected SCF for molecular interactions.\cite{stoll1980use, gianinetti1996modification}
This strategy enables the straightforward calculation of energies and properties by expanding the MOs of each fragment in terms of the atomic orbitals of the same fragment, resulting in orthonormal MOs for each fragment (whereas MO sets belonging to different fragments are not mutually orthonormal). 
Charge transfer, where the MOs of the acceptor fragment accommodate part of the electron density from the donor fragment, cannot be represented.
In contrast, the transfer of integer charges --- \textit{i.e.} a full electron in the case of ICT --- can be computed straightforwardly by treating the donor and acceptor monomers as cation and anion, respectively. 
In addition to the ALMO method, where we keep the above constraint during the variational optimization, we also evaluate the performance of the IMOM and SGM algorithms on the guess MOs provided by the ALMO method for the tetrafluoroethylene-ethylene model system in section \ref{sec:tfetet}. \\
Finally, in section \ref{sec:clevercage} we compare OO-DFT methods to reparametrizations of the LRC-$\omega$PBE\cite{rohrdanz2008simultaneous} XCF in TD-DFT.
A system-specific tuning of the range-separation parameter $\omega$ proved effective in combination with TD-DFT by recovering properties of the exact XCF with semi-empirical RSH XCFs\cite{prokopiou2022optimal}.
Two alternative schemes to determine a suitable $\omega$ in LRC-$\omega$PBE have recently been proposed. 
The global density-dependent (GDD) tuning\cite{modrzejewski2013density, mandal2025simplified} adjusts $\omega$ by computing the distance $d_\mathbf{x}$ between an electron in the outer regions of a molecule and the exchange hole in the localized valence orbitals
\begin{align}
\omega_{\mathrm{GDD}}=C\langle d_{\mathbf{x}}^{2}\rangle^{-1/2},
\end{align}
where $C$ is an empirical constant determined for each XCF.
Automatic GDD tuning is implemented in Q-Chem via ground-state density calculation for a system of interest computed with an initial guess for $\omega$ to determine $d_\mathbf{x}$.
The optimal $\omega_\mathrm{GDD}$ can then be directly plugged into the expression for the LRC-$\omega$PBE XCF to be used in subsequent calculations.
GDD tuning is size-extensive and has successfully been applied to the calculation of intramolecular charge-transfer excitations and extended aromatic systems.\cite{mandal2025simplified}.
In contrast, a common problem in alternative tuning strategies comes from rather abrupt changes in the optimal range-separation parameter for similar systems, \textit{e.g.} different conformations of the same chemical compound, which results in discontinuous properties. \\
Yan \textit{et al.} proposed an alternative empirical XCF reparametrization\cite{} designed for the description of CT states. 
From a preliminary TD-DFT calculation using a related XCF (\textit{e.g.} the PBE0\cite{adamo1999toward} XCF is used for tuning LRC-$\omega$PBE\cite{rohrdanz2008simultaneous}), and the average excitation distance $\mathrm{D}_\mathrm{CT}$ is computed for several CT excited states, respectively.
The $\omega$ parameter for the final calculations is then obtained from the maximum $\mathrm{D}_\mathrm{CT}$ in the set
\begin{align}
\omega^* = \frac{2}{\mathrm{D}_\mathrm{CT}^\mathrm{max}},
\end{align}
where the 2 in the numerator is an empirical constant.
The calculation of absorption spectra in push-pull organic dyads with LRC-$\omega^*$PBE showed promising results\cite{yan2025adaptable}, improving intramolecular CT excitations while preserving the accurate prediction of valence states.

\section{Computational methods}

All OO-DFT and most TDA calculations presented in this work are performed with a development version of Q-Chem\cite{epifanovsky2021software} using the def2-TZVP\cite{weigend2005balanced} basis set. Calculations on the tetrafluoroethylene-ethylene model system from section \ref{sec:tfetet} employ the LRC-$\omega$PBE\cite{rohrdanz2008simultaneous} XCF. The guess refinement method introduced in section \ref{sec:frzopt} is combined with the IMOM method\cite{barca2018simple} and the SGM algorithm\cite{hait2020excited, hait2021orbital}, performing a line search at each iteration. The ICT excitations in the large PTZ-ANQ molecular dimers in section \ref{sec:clevercage} are computed with TDA using two alternative reparametrizations of the LRC-$\omega$PBE XCF, where the range-separation parameter $\omega$ is adjusted according to properties of the exchange hole\cite{modrzejewski2013density, mandal2025simplified}, or the excited-state density\cite{yan2025adaptable}. Corresponding OO-DFT calculations with the ALMO-SGM and FR-SGM methods are performed with the $\omega$B97X-D XCF\cite{chai2008long}, and the def2-ECP effective core potential\cite{grimme2015consistent, sure2013corrected} is used for the calculation of the Pd ions in the coordination cage structure. 
In section \ref{sec:dye-TiO2}, we compute the low-lying excitations in the two dye-TiO$_2$ complexes with the simplified TDA\cite{grimme2013simplified} (sTDA) method implemented in Orca\cite{neese2022software}, and recompute the corresponding excitations with the FR-SGM method, with the $\omega$B97X-D\cite{chai2008long} XCF.

\section{Results and discussion}

\subsection{Constrained optimization algorithm}\label{sec:frzopt}

\begin{figure}[h!]
 \centering
\includegraphics[width=0.85\textwidth]{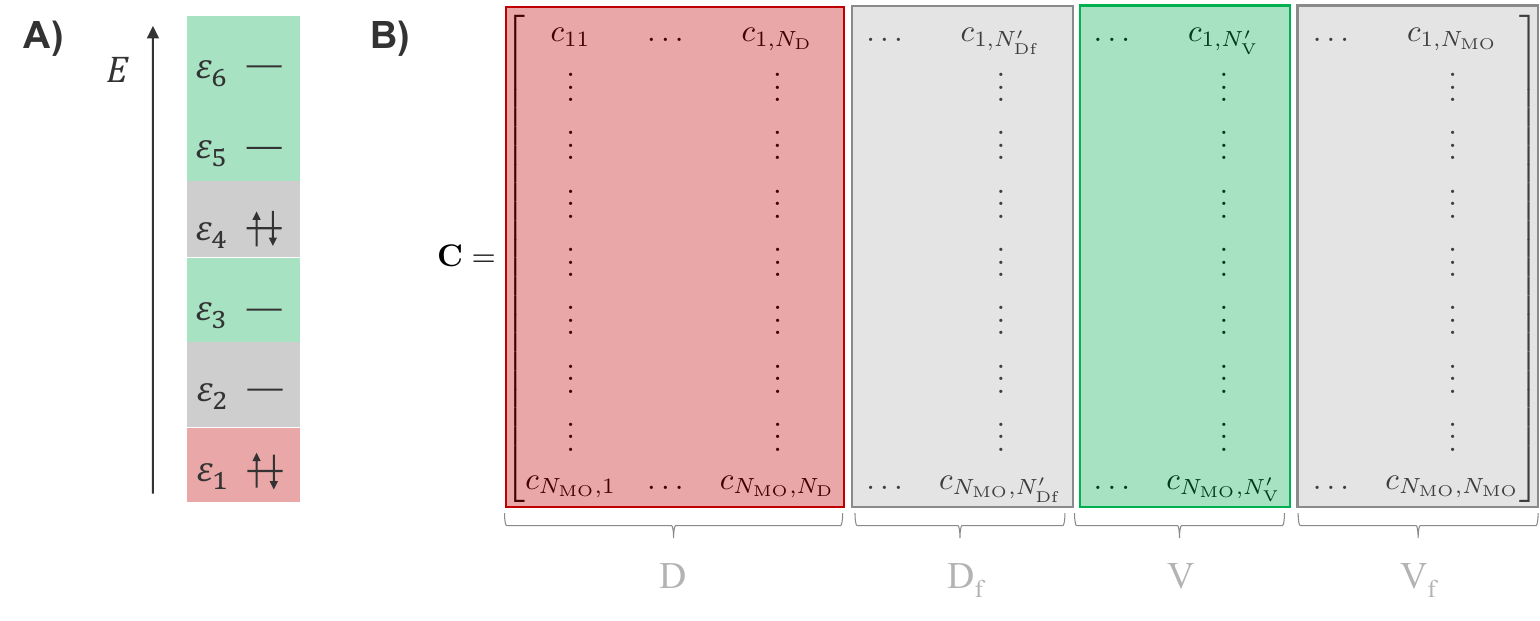}
 \caption{$\mathbf{C}$ matrix partitioning for frozen electron-and-hole constrained optimization. Panel A shows the example of a double excitation from MO no. 2 to MO no. 4, whereas the $\mathbf{C}$ matrix is given in the general form in panel B. Free doubly-occupied vectors are highlighted in red, free virtual orbitals in green, and all frozen vectors in grey. A detailed discussion of the index notation used in panel B and the $\mathbf{C}$ matrix reordering procedure is given in the main text.}
 \label{fgr:Cmat-new}
\end{figure}

ICT excitations involve major changes in the electron density because, ultimately, an ion pair is formed. 
The resulting Coulombic attraction changes the electrostatic problem substantially, and any initial guess that takes this into account should be a good starting point for OO-DFT calculations of CT states.
The ALMO method we described in the previous section achieves this if the fragments are chosen in their respective ionic states.
An alternative initial guess can be generated by freezing the electron and hole orbitals during a variational optimization as suggested by Schmerwitz \textit{et al.}\cite{van2002geometric}.
We implemented this initial guess by rearranging the molecular orbital coefficient matrix $\mathbf{C}$, thereby effectively modifying the Aufbau principle and subsequently excluding the occupied-virtual rotations of the frozen electron and hole orbitals from the electronic gradient.
Hence, the resulting calculation is a constrained minimization including all occupied-virtual rotations for MOs that are not directly involved in the excitation. \\
We start from a partitioning of the orbital coefficient matrix $\mathbf{C}$  of a restricted Hartree--Fock (RHF) calculation but note that this approach can be extended straightforwardly to the unrestricted case and Kohn--Sham theory.
All $\mathbf{C}$ elements are real numbers and $\mathbf{C}$ is ordered according to the block structure shown in  Fig.~\ref{fgr:Cmat-new}. 

\begin{figure}[h]
 \centering
 \includegraphics[width=0.95\textwidth]{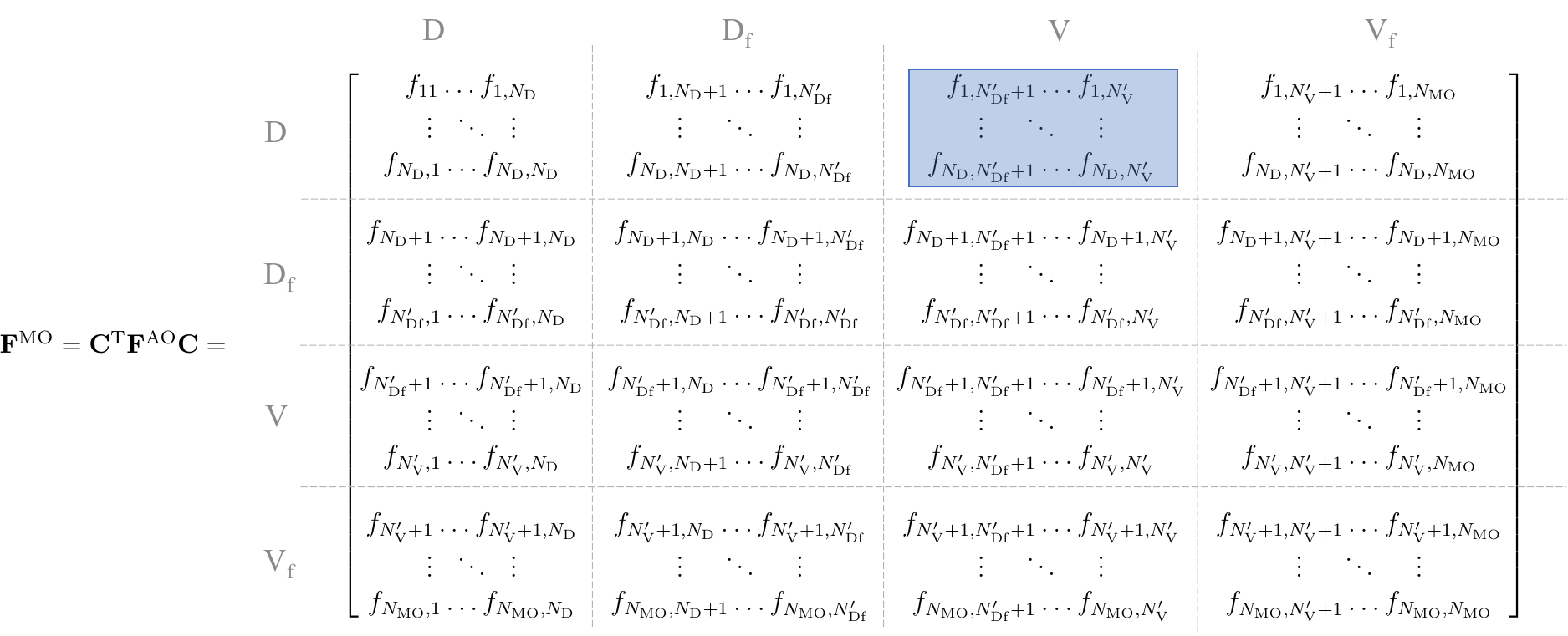}
 \caption{$\mathbf{F}$ matrix block structure arising from the $\mathbf{C}$ matrix partitioning introduced in Fig. \ref{fgr:Fmat}. 
 The Fock matrix has 16 blocks, six in the upper triangle. Frozen blocks do not contribute to the gradient, such that only the DV block from the upper triangle (highlighted in blue) contributes to the gradient.}
 \label{fgr:Fmat}
\end{figure}

Given an atomic orbital (AO) basis of size $N_\mathrm{basis}$, the $\mathbf{C}$ matrix of dimension $N_\mathrm{MO}\ (\leq N_\mathrm{basis})$ is ordered into blocks. ``Free'' doubly-occupied vectors are in the first block D, indexed from 1 to $N_\mathrm{D}$. 
A block D$_\mathrm{f}$ with $N_\mathrm{Df}$ frozen doubly-occupied orbitals follows, indexed from $N_\mathrm{D}+1$ to $N_\mathrm{D}+N_\mathrm{Df}=N'_\mathrm{Df}$. 
Similarly, virtual orbitals are partitioned into the V block containing free virtual orbitals ($N'_\mathrm{Df}+1, ..., N'_\mathrm{Df}+N_\mathrm{V}=N'_\mathrm{V}$) and a block  V$_\mathrm{f}$ containing frozen virtual orbitals ($N'_\mathrm{V}+N_\mathrm{Vf}+1, ..., N_\mathrm{MOs}$). 
Note that for ICT excitations the D$_\mathrm{f}$ and N$_\mathrm{f}$ blocks are simple column vectors.
At the beginning of the calculation, occupied (red) and virtual (green) orbitals are reordered to move the frozen vectors (in grey) to the end of each block. 
In the simple example shown in panel A of Fig.~\ref{fgr:Cmat-new}, a double excitation is computed by exciting one $\alpha$ and one $\beta$ electron from MO no. 2 to MO no. 4, and the MOs are sorted in the order 1-4-3-5-6-2. The D$_\mathrm{f}$ and V$_\mathrm{f}$ blocks contain only MOs no. 4 and 2, respectively.
The block structure emerging for the $\mathbf{C}$ matrix by partitioning into active and frozen parts is depicted in panel B of Fig.~\ref{fgr:Cmat-new}, using the general notation introduced above.
Once the matrix is reordered to have the MOs involved in the excitation at the end of the occupied and virtual block, the constrained optimization just requires input on the number of holes and electrons to constrain, and in principle, multiple electron-hole pairs can be considered. \\
The energy optimization is then performed with a second-order method, which requires gradients and previous steps to compute an approximation of the inverse Hessian. 
In our implementation, we leverage the efficient GDM algorithm\cite{van2002geometric} implemented in Q-Chem. 
The $\mathbf{C}$ matrix reordering results in the partitioning of the Fock matrix depicted in Fig.~\ref{fgr:Fmat}. The Fock matrix is computed in the MO basis.
The energy gradient with respect to the MO rotations is given by the off-diagonal elements of the symmetric Fock matrix. 
In the presence of frozen blocks D$_\mathrm{f}$ and V$_\mathrm{f}$, the Fock matrix is partitioned into 16 blocks, six belonging to the upper triangle. 
Only the DV block in the upper triangle (highlighted in blue in Fig. \ref{fgr:Fmat} enters the gradient, as frozen vectors do not contribute in this constrained optimization. 
Similarly, gradients and steps from previous iterations are stored only for the free vectors and are transformed by taking the corresponding blocks from the transformation matrix given the step $\Delta$. 
We evaluate the performance of this refined guess strategy (termed FR) combined with the IMOM and SGM algorithms on the lowest-lying ICT excitation of the tetrafluoroethylene-ethylene dimer at various donor-acceptor displacements in the following section \ref{sec:tfetet}.

\subsection{Convergence study: tetrafluoroethylene-ethylene dimer}
\label{sec:tfetet}

\begin{figure}[h!]
\centering
\includegraphics[width=0.75\textwidth]{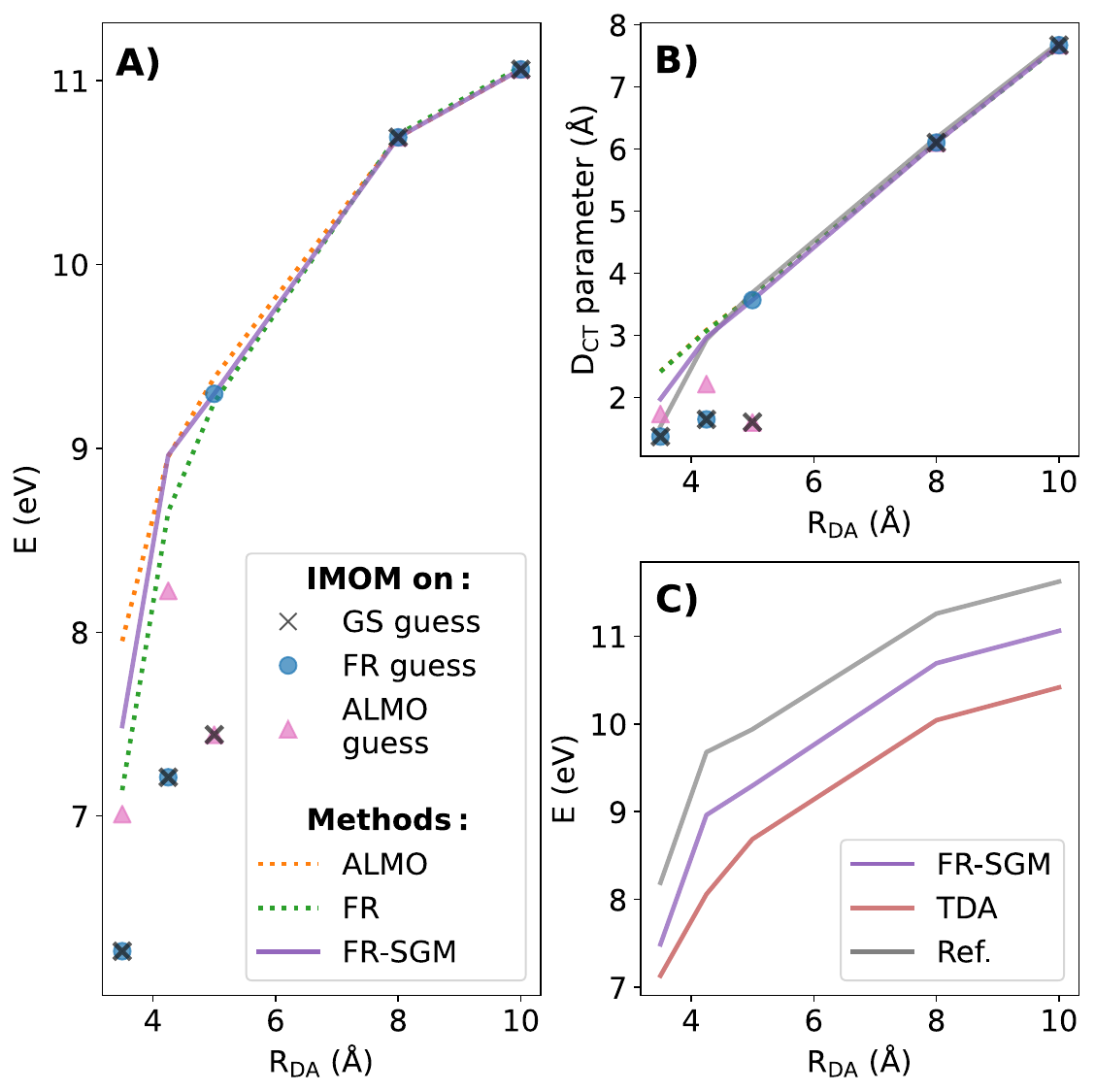}
  \caption{Lowest-lying ICT excitation in the tetrafluoroethylene-ethylene dimer as a function of donor-acceptor distance $R_\text{DA}$.
  The convergence of several OO-DFT methods and initial guess MOs is analyzed in panel A, and the accuracy of the FR-SGM method is compared against TDA and the EOM-CCSD(fT) reference in panel C.
  The density-based CT descriptor $\mathrm{D}_\mathrm{CT}$ is plotted in panel B.
  The markers show results for IMOM calculations computed with the ground-state MO guess orbitals (grey x markers), ALMO guess orbitals (pink triangles), and frozen hole-and-electron guess orbitals (blue dots).
  The dashed orange lines correspond to the energy and descriptors computed on the ALMO density, and the dashed green lines belong to the frozen hole-and-electron density.
  The purple line is obtained by using the FR guess orbitals to initiate the SGM algorithm with line search (FR-SGM).
  In panel C, the red line is the TDA result, and the reference (grey line) is computed with EOM-CCSD(fT) and the cc-pVTZ basis.}
  \label{fgr:convergence-new}
\end{figure}

In our previous work on ICT excitations,\cite{D4CP01866D} we observed that especially for short donor-acceptor distances $R_\mathrm{DA}$=3.5-5 \r{A}, convergence to the desired ICT states is unreliable and variational collapse to different excited states or even the ground state occurs frequently.
In contrast, Schmerwitz\cite{schmerwitz2025freeze} \textit{et al.} demonstrated that for the example of the tetrafluoroethylene-ethylene dimer system, their freeze-and-release algorithm reliably converges to the lowest ICT state.
Here, we analyze how far this is due to an improved initial guess rather than the optimization algorithm. 
We therefore combined initial guesses provided by the FR method presented in the previous section and the ALMO guess with IMOM and SGM optimization for the same dimer system.
In our calculations, IMOM is coupled to the DIIS extrapolation algorithm, and the SGM algorithm performs a line search at each iteration. All results on the tetrafluoroethylene-ethylene dimer are converged to $10^{-5}$ H.
We measure the ICT character of the states by the average electron-hole distance parameter $\mathrm{D}_\mathrm{CT}$ (panel B of Fig.~\ref{fgr:convergence-new}), which can be readily computed from a grid representation of the ground- and excited-state electron densities and were introduced by Le Bahers \textit{et al.} in Ref.~\citenum{le2011qualitative}.\\

The energies of the lowest ICT states as a function of the donor-acceptor distance are shown in panel A of Fig.~\ref{fgr:convergence-new}. 
The orange and green dotted lines result from calculations where the ALMO and FR constraints are sustained, respectively, whereas the purple line corresponds to a subsequent variational optimization with SGM (FR-SGM). We note that starting SGM from the ALMO guess yields identical results in this example. 
For large donor-acceptor distances, all methods presented here reliably converge to an ICT state that shows the expected $1/R_\text{DA}$ behaviour.
Evidently, even at short distances, the ALMO and FR constraints provide excellent guesses that are only minimally optimized during the subsequent unconstrained variational optimization.
At long donor-acceptor distances, this discrepancy is invisible on the scale of the graph.
The CT descriptor in panel B confirms convergence to an ICT state.\\
Different results are obtained when these initial guesses are combined with a subsequent IMOM optimization.
Starting from a ground-state (gray x-markers), FR guess (blue dots), or ALMO (pink triangles), panel A of Fig.~\ref{fgr:convergence-new} shows that for short donor-acceptor distances, IMOM optimization converges to states that are significantly lower in energy than the initial guesses or the SGM results described above.
Predictably, the discrepancy is worst on average for the ground state guess, but also the refined guess strategies cannot guarantee convergence to the targeted ICT state.
Indeed, and in line with our previous work,\cite{D4CP01866D} panel B demonstrates that these calculations converged to lower-lying non-ICT states.\\
We hence conclude that ALMO and FR provide excellent initial guesses for ICT states, and SGM provides a reliable convergence to the targeted states once the constraints are lifted in a variational optimization.

We further compare the OO-DFT results to TD-DFT calculations (red line) and an accurate equations-of-motion coupled-cluster (EOM-CC) reference calculation (gray line) in panel C of Fig.~\ref{fgr:convergence-new}.
The reference densities (panel B in Fig.~\ref{fgr:convergence-new}) were computed with EOM-CCSD, and the energies (panel C) were adjusted by perturbative triples correction with the EOM-CCSD(fT) method.
All DFT methods are red-shifted by an offset, but as Table~\ref{tbl:accuracy} shows, the OO-DFT methods yield a much smaller mean error (0.64~eV) than TDA (1.27~eV). 
The improvement achieved by the OO-DFT method is attributed to the accuracy of the range-separated hybrid XCF, combined with an algorithm that ensures convergence to the targeted state.
The approximation of weakly interacting fragments introduced in the ALMO method is invalid for short donor-acceptor separations, such that the stabilization achieved upon variational optimization for this situation with significant orbital overlap between the fragments is not captured. 
The statistical errors for the constrained ALMO results can hence look better than the results of the OO-DFT methods (here: FR-SGM) but the former does not correspond to saddle points on the electronic hypersurface and are therefore not valid excited states.
The TDA method gives the largest mean error and mean variance in Table \ref{tbl:accuracy}.
We observe, however, a very small variance for FR-SGM, meaning that the energies deviate from the reference energies by a rather constant shift.
A systematic offset can be beneficial compared to a statistical error in the prediction of deactivation paths if all excited states are affected by the same shift, resulting in fortuitous error compensation.
Interestingly, the FR energy (green dotted line) lies lower than the target stationary points at short distances, indicating that the saddle-point search must proceed uphill along the electron-hole MO rotation, which is possible since we optimize the squared gradient.
Furthermore, the $\mathrm{D}_\mathrm{CT}$ descriptor (panel B) shows how FR-SGM is the only method correctly describing the electron-hole interaction at 3.5 \r{A} separation, deviating from the linear trend just as the reference method. \\
We conclude that when started with a suitable initial guess, the SGM method converges reliably to the desired stationary point on the excited-state surface. In combination with a suitable range-separated hybrid XCF, OO-DFT methods yield a small mean error and variance with respect to highly accurate reference calculations.

\begin{table}[H]
\small
  \caption{Mean signed error (MSE) and mean signed variance (MSV) on the
ICT excitation energy for the tetrafluoroethylene-ethylene dimer system $R_\mathrm{DA}$ scan for each excited-state DFT method.
All calculations are performed with the LRC-$\omega$PBE XCF and the def2-TZVP basis. All values reported in eV unit}
  \label{tbl:accuracy}
  \begin{tabular*}{0.68\textwidth}{@{\extracolsep{\fill}}lll}
    \hline
Method &  Mean error (eV) &  Mean variance (eV)\\
    \hline
   TDA &   -1.27 &       0.043 \\
  ALMO &   -0.53 &       0.031 \\
    FR &   -0.77 &       0.062 \\
FR-SGM &   -0.64 &       0.005 \\
    \hline
  \end{tabular*}
\end{table}

\subsection{Charge-transfer excitations in a donor-acceptor Pd coordination cage}
\label{sec:clevercage}

\begin{figure}[h!]
\centering
\includegraphics[width=0.9\textwidth]{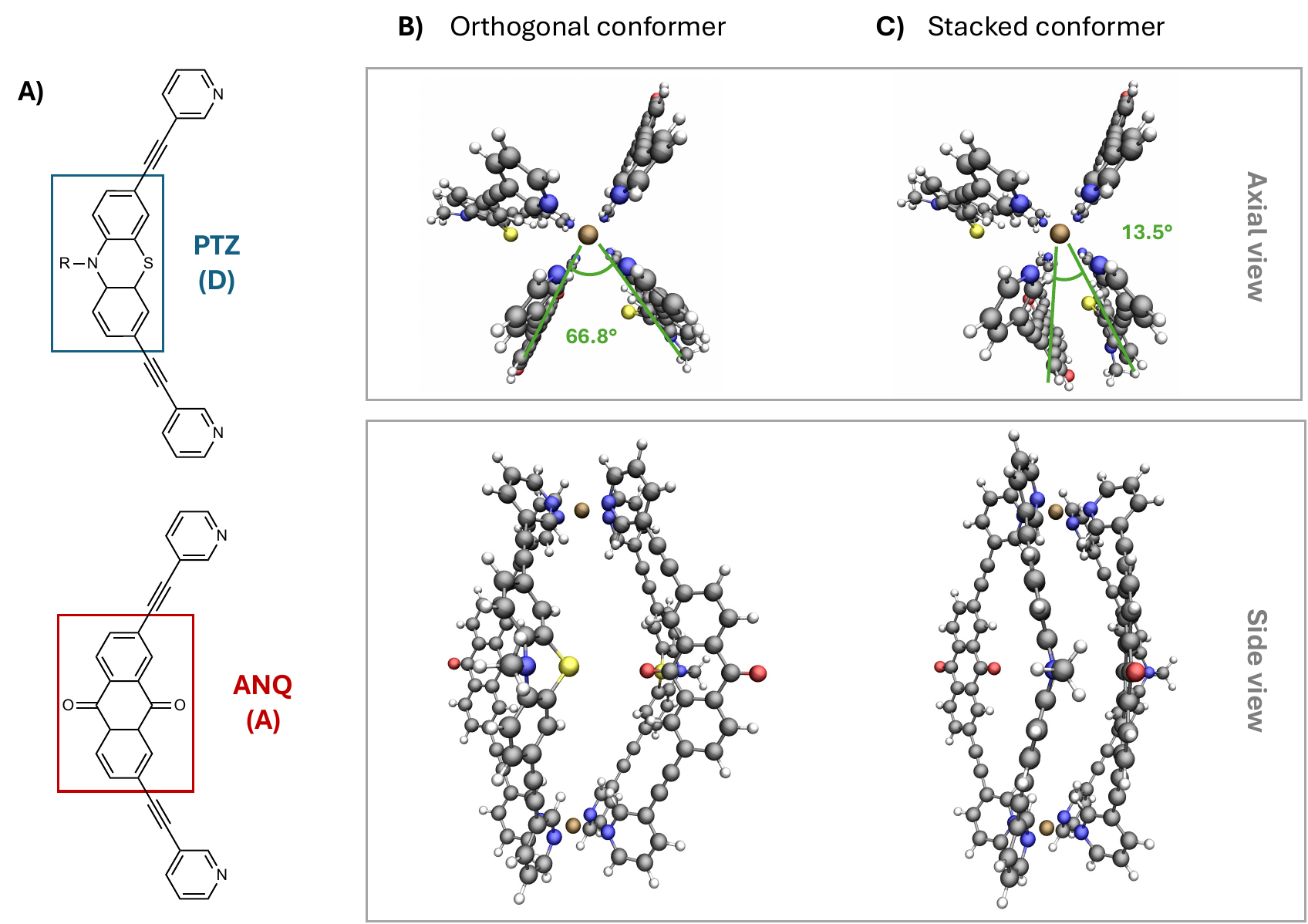}
  \caption{ Panel A shows the structures of the dye-bearing ligands with dye cores highlighted in both the donor (D) and acceptor (A) ligand.
  Panels B and C: Stable [Pd$_2$(PTZ)$_2$(ANQ)$_2$]$^{4+}_\text{trans}$ coordination cage conformations obtained by optimizations with the $\omega$B97X-D XCF, def2-SVP basis set, and def2-ECP effective core potentials.
  Optimization of an initial guess with 90$^\circ$ angles between the ligands leads to the orthogonal conformer in panel B, whereas a stable stacked conformer is obtained by reducing the D-A dihedral and shown in panel C.
  All structures are displayed along the Pd-Pd axis (axial view, top) and from the side (bottom).}
  \label{fgr:conformations-new}
\end{figure}

\begin{figure}[H]
\centering
  \includegraphics[width=\textwidth]{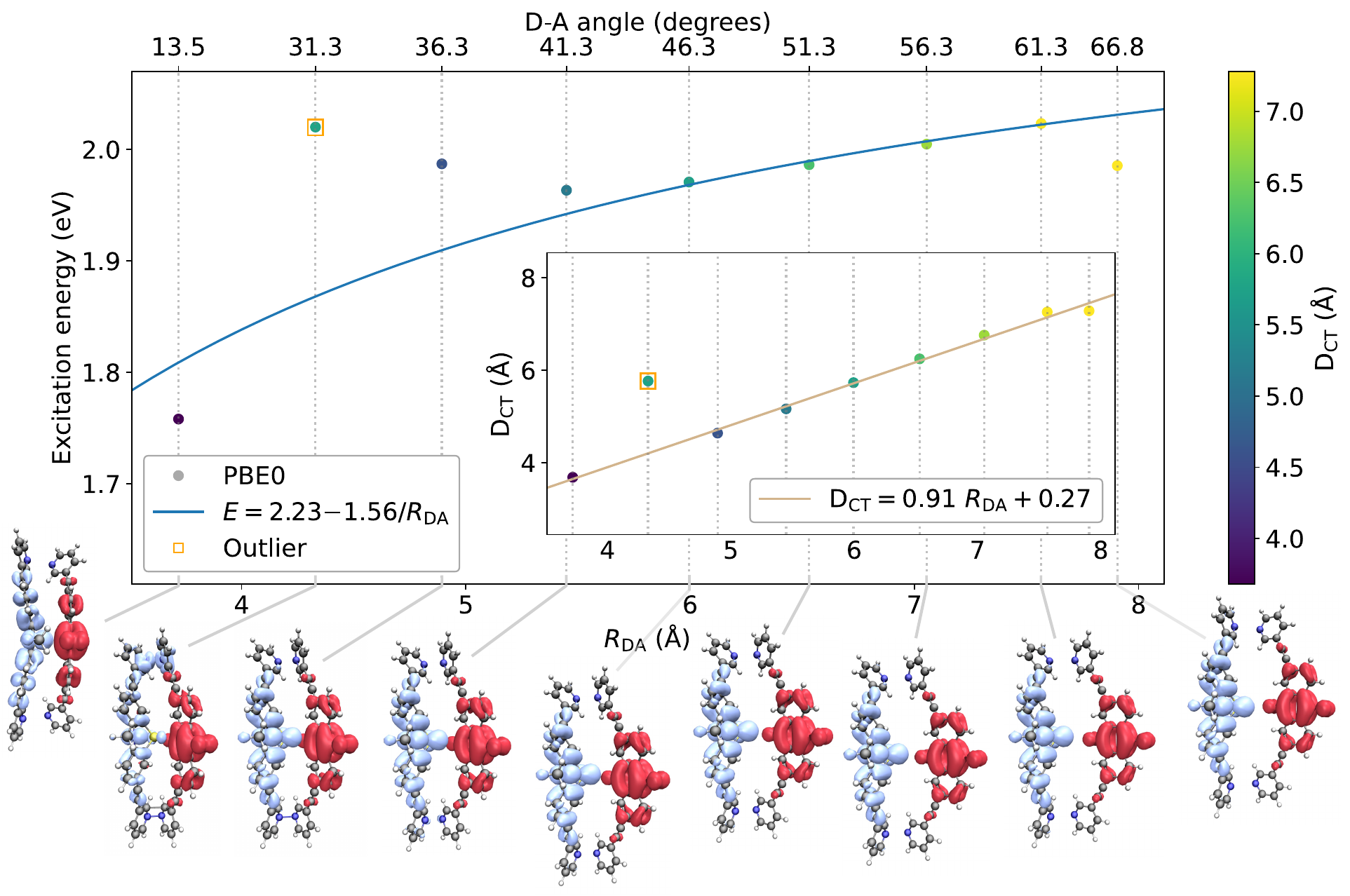}
  \caption{Lowest-lying ICT excitation energies of several structures with varying dihedral angle between donor and acceptor obtained by interpolating the orthogonal and stacked structures.
  The density difference between the ground and ICT excited state is shown at the bottom as an isosurface plot (isovalue 0.0005 for the blue surface and -0.0005 for the red surface).
  The value of $\mathrm{D}_\mathrm{CT}$ is plotted in the inset and is used to color-code the scatter plots.
  The energies are fitted with a horizontal asymptote, and the $\mathrm{D}_\mathrm{CT}$ plot is fitted with a linear expression.
  Data points marked with an orange square are identified as outliers and excluded from both fits.}
  \label{fgr:pbe0}
\end{figure}

Since our goal is to calculate ICT in large supramolecular systems, we require scalable excited-state electronic-structure methods and identified FR-SGM as a suitable approach.
In this section, we focus on a supramolecular cage structure, where four banana-shaped ligands coordinate two square-planar Pd ions.\cite{han2014self}
The Clever group pioneered the design of such supramolecular coordination cage complexes by integrating dyes into the structural elements of the supramolecule, and investigated an interlocked double-cage with electron-rich phenothiazine (PTZ) and electron-deficient anthraquinone (ANQ) dyes in the walls. \cite{frank2016light}
The ligands used for the preparation of the Pd cage are displayed in Fig.~\ref{fgr:conformations-new} (panel A, note that the R residue in the PTZ ligand is a methyl group in our calculations).
The formation of a PTZ-ANQ charge-transfer excitation upon irradiation with a 400 nm photon was investigated by time-resolved spectroscopy and spectro-electrochemistry.
We consider this a very relevant system to compute with scalable electronic-structure methods, due to the presence of multiple CT excitations involving the organic dyes and the metal ions, which are affected by changes in the cage conformation.
For simplicity, we do not compute the interlocked double-cage structure and focus on a single coordination cage with two PTZ and two ANQ dyes in trans conformation, [Pd$_2$(PTZ)$_2$(ANQ)$_2$]$^{4+}_\text{trans}$.
The PTZ and ANQ ligands were built using Avogadro\cite{hanwell2012avogadro} 1.2 and were positioned on perpendicular planes to saturate the coordination sphere of two square-planar Pd ions.
In the following, we refer to PTZ as the ``donor'' ligand D and ANQ as the ``acceptor'' ligand A.
The cage structure was optimized with $\omega$B97X-D in a def2-SVP basis, and def2-ECP effective core potentials.
Optimization of the ``orthogonal'' guess structure relaxed to a local minimum, with well-separated dyes (Fig.~\ref{fgr:conformations-new}, panel B).
An alternative conformation (Fig.~\ref{fgr:conformations-new}, panel C) was prepared by slightly tilting the dihedral angle defined by the dye planes and reoptimizing the structure, where the PTZ and ANQ ligands participated in $\pi-\pi$ stacking interaction.
The electronic energy of the stacked and orthogonal conformers differs by 0.61~kcal/mol, so we assume both conformers to be accessible at room temperature.
In the first section of our results, TD-DFT and OO-DFT methods are compared by computing the charge-transfer excitations in the D-A dimer extracted from each cage conformation, neglecting the Pd ions and the dye pair not involved in the excitation.
Additional conformations interpolating the orthogonal and stacked forms were constructed by manipulation of the Z-matrix using Molden,\cite{schaftenaar2000molden} producing seven intermediate conformations, for a total of nine conformers.
The D-A dihedral angles are measured by fitting a plane through each dye; since the PTZ ligand is not fully planar, the measured dihedral is 66.8\degree\,  in the orthogonal conformer, whereas the stacked conformer is characterized by a D-A dihedral of 13.5\degree.
The dihedral angles for the intermediates measured in this way are 31.3, 36.3, 41.3, 46.3, 51.3, 56.3, and 61.3\degree.
In the second section of the results, low-lying charge-transfer excitations of the whole Pd cage system are computed for a set of interpolated structures connecting the stacked and orthogonal conformers.
These interpolated structures were obtained using the Atomistic Simulation Environment.\cite{larsen2017atomic, smidstrup2014improved}, employing the Image-Dependent Pair Potential\cite{smidstrup2014improved} (IDPP) algorithm.\\
Fig.~\ref{fgr:pbe0} shows the TDA results obtained with the PBE0\cite{adamo1999toward} XCF for the lowest-lying ICT excitation of the D-A dimer in the conformations of the PTZ-ANQ cage, as a function of the DA dihedral angle (and corresponding distance) discussed above.
We define the donor-acceptor distance $R_\text{DA}$ as the distance between nuclei of the dye centers highlighted in panel A of Fig.~\ref{fgr:conformations-new}.
The $\mathrm{D}_\mathrm{CT}$ magnitude is color-coded, and the $\mathrm{D}_\mathrm{CT}$ is additionally shown separately in the inset.
The excitation energy is fitted with the asymptotic expression
\begin{align}
E=a-\frac{b}{R_\mathrm{DA}}\, ,
\label{eq:asy}
\end{align}
where $a$ and $b$ are fitting parameters, and $a$ can be identified with $IP_\text{D} - EA_\text{A}$, according to Mulliken's formula.\cite{mulliken1952molecular}
$\mathrm{D}_\mathrm{CT}$ is fitted linearly as in
\begin{align}
\mathrm{D}_\mathrm{CT}=k R_\mathrm{DA} + q \, ,
\end{align}
where $k$ and $q$ are fitting parameters.

The fitted parameters are reported in the respective Figures.
Data points marked with an orange square are outliers excluded from both fits.
An isosurface plot of the density difference w.r.t. the electronic ground state is displayed at the bottom of Fig.~\ref{fgr:pbe0} for all conformations, using an isovalue -0.0005 for the red surface and isovalue 0.0005 for the blue surface.
Overall, the results obtained with PBE0 demonstrate that the global hybrid XCF captures only a fraction of the hole-electron attraction, blue-shifting the ICT excitation energy by only 0.2 eV over the dihedral scan.
The fit shows that $a=I P_{\mathrm{D}}-E A_{\mathrm{A}}$ is significantly underestimated compared to the more accurate methods discussed below, which is expected from a global-hybrid XCF approximation\cite{janesko2009screened, sohlberg2020s}.
Values for $I P_{\mathrm{D}}-E A_{\mathrm{A}}$ are reported in Table~\ref{tbl:ipea} for the DFT methods investigated here, alongside the values for $b$ in equation~\ref{eq:asy}. The incomplete description of the electron-hole attraction by the global hybrid XCF affects all data points in the $R_\mathrm{DA}$ scan, and the resulting curve is not as steep as other methods, evident by the smaller $b$.

\begin{table}[h!]
\small
  \caption{Results of the fit over the energy expression in Eq.~\ref{eq:asy} for the lowest-lying ICT excitation in the PTZ-ANQ dimer over the D-A dihedral scan for different methods and XCFs.}
  \label{tbl:ipea}
  \begin{tabular*}{0.78\textwidth}{@{\extracolsep{\fill}}llll}
    \hline
  Method &  XCF & $a=I P_{\mathrm{D}}-E A_{\mathrm{A}}$ (eV) & $b$ (eV/\r{A}) \\
    \hline
   TDA   &  PBE0\cite{adamo1999toward}  &  2.23 &  1.56 \\
   TDA   &  LRC-$\omega_\mathrm{GDD}$PBE\cite{modrzejewski2013density, mandal2025simplified}  &  4.64 &  8.02 \\
   TDA   &  LRC-$(2/\mathrm{D}_\mathrm{CT})$PBE\cite{yan2025adaptable}  &  5.38 & 15.11 \\
   ALMO-SGM &  $\omega$B97X-D\cite{chai2008long}  &  4.18 &  6.24 \\
   FR-SGM   &  $\omega$B97X-D\cite{chai2008long}  &  4.22 &  5.87\\
    \hline
  \end{tabular*}
\end{table}

\begin{figure}[h!]
 \centering
 \includegraphics[width=\textwidth]{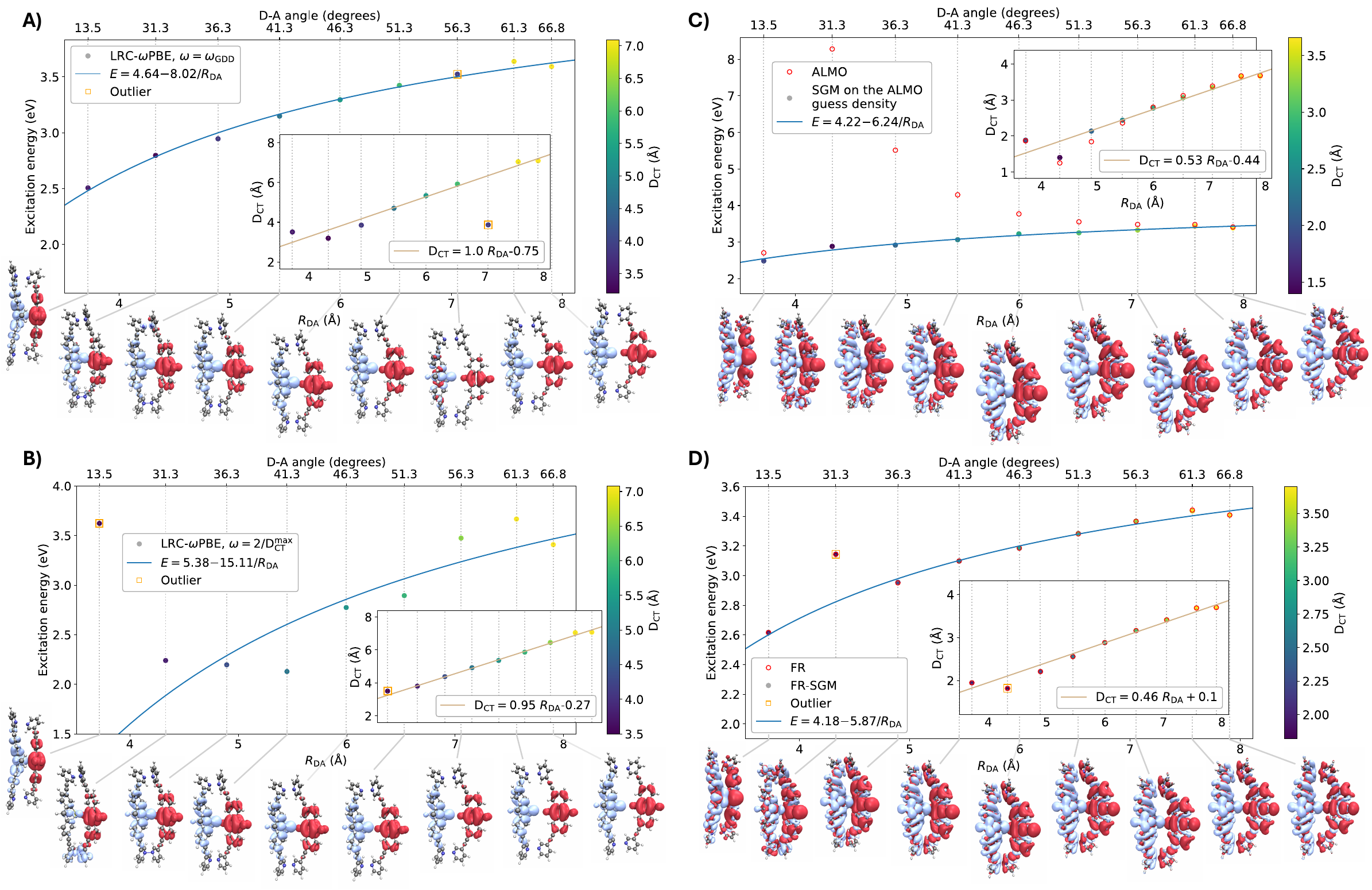}
 \caption{Lowest-lying ICT excitation energies of several structures with varying dihedral angle between donor and acceptor obtained by interpolating the orthogonal and stacked structures.
  The density difference between the ground and ICT excited state is shown at the bottom of each panel as an isosurface plot (isovalue 0.0005 for the blue surface and -0.0005 for the red surface).
 The ICT excitation energies are  computed with TDA, using $\omega_\mathrm{GDD}$ or a $\mathrm{D}_\mathrm{CT}$-based tuning in panels A and B, respectively.
 The results from the OO-DFT methods ALMO-SGM and FR-SGM are shown in panels C and D, respectively.
 In the plots of the OO-DFT methods, the initial guess energy and $\mathrm{D}_\mathrm{CT}$ are shown as red circles for each structure.
 The value of $\mathrm{D}_\mathrm{CT}$ is plotted in the inset and is used to color-code the scatter plots.
 The energy plot is fitted with a horizontal asymptote, and the $\mathrm{D}_\mathrm{CT}$ plot is fitted with a linear expression.
 Data points marked with an orange square are identified as outliers and excluded from both fits.}
 \label{fgr:DAscans}
\end{figure}

Fig.~\ref{fgr:DAscans} shows the results of the dihedral scan for the remaining electronic-structure methods in Table~\ref{tbl:ipea}.  
Panel A is obtained by applying the GDD reparametrization scheme, tuning $\omega_\mathrm{GDD}$ for each conformation in the scan.
The resulting curve is smooth, with an asymptote that is blue-shifted by more than 2~eV compared to the non-range-separated XCF. The data point corresponding to 56.3\degree~dihedral is excluded from the fit, since the calculated D$_\text{CT}$ value strongly deviates from the expected linear trend.

Results obtained from the empirical parametrization\cite{yan2025adaptable} suggested by Yan \textit{et al.} are shown in panel B of Fig.~\ref{fgr:DAscans}.
The 40 lowest-lying singlet excited states are computed with the PBE0 XCF (results reported in Fig.~\ref{fgr:pbe0}), and the state with the maximum $\mathrm{D}_\mathrm{CT}$ is used to compute $\omega^*$.
The resulting curve is steeper than for the previous results and results in the largest $I P_{\mathrm{D}}-E A_{\mathrm{A}}$ and $b$ obtained for all methods.
The data points corresponding to the stacked conformer were removed from the fit, but the obtained energies scatter slightly around the fitted trend.
To better understand the reason for the smoother curve obtained with the $\omega_\mathrm{GDD}$-tuning procedure as compared to the tuning based on the maximum $\mathrm{D}_\mathrm{CT}$, we report the respective range-separation parameters in Table~\ref{tbl:omegas}.
The negligible variation of $\omega_\mathrm{GDD}$ along the scan explains the smooth and monotonous curve obtained with this tuning procedure.
On the contrary, major changes of $\mathrm{D}_\mathrm{CT}^\mathrm{max}$ result in strong variations of $\omega^*$ for neighboring data points in the scan.
Notably, for large donor-acceptor separations, the results are almost identical.
Future work might reveal if empirical tuning based on the $\mathrm{D}_\mathrm{CT}$ descriptor in the asymptotic limit can generally improve the results. 
In conclusion, for systems where conformational changes can significantly affect the energy of CT excitations, size-consistent reparametrizations like $\omega_\mathrm{GDD}$ are recommended for accurate TD-DFT energetics of ICT states.

\begin{table}[h!]
\small
\caption{$\omega^\mathrm{GDD}$\cite{modrzejewski2013density} values obtained applying the general density-dependent reparametrization, $\omega^*$ obtained using the parametrization introduced by Yan \textit{et al.}\cite{yan2025adaptable}, and $\mathrm{D}_\mathrm{CT}^\mathrm{max}$ over the 40 lowest-lying roots computed with the PBE0 XCF 
  for the ANQ-PTZ dimer for the same conformations as in Fig.~\ref{fgr:pbe0} and \ref{fgr:DAscans}.}
  \label{tbl:omegas}
  \begin{tabular*}{0.78\textwidth}{@{\extracolsep{\fill}}llll}
    \hline
D-A dihedral $\theta$ (\degree) & $\omega_\mathrm{GDD}$ (\r{A}$^{-1}$) & $\omega^*=2/\mathrm{D}_\mathrm{CT}^\mathrm{max}$ (\r{A}$^{-1}$) & $\mathrm{D}_\mathrm{CT}^\mathrm{max}$ (\r{A}) \\
    \hline
13.5 & 0.242 & 0.467 &  4.28 \\
31.3 & 0.240 & 0.152 & 13.12 \\
36.3 & 0.240 & 0.153 & 12.82 \\
41.3 & 0.242 & 0.142 & 14.12 \\
46.3 & 0.239 & 0.188 & 10.63 \\
51.3 & 0.237 & 0.187 & 10.72 \\
56.3 & 0.235 & 0.229 &  8.74 \\
61.3 & 0.234 & 0.237 &  8.45 \\
66.8 & 0.233 & 0.215 &  9.31 \\
    \hline
  \end{tabular*}
\end{table}

Finally, we also investigated the performance of the ALMO-SGM and FR-SGM OO-DFT methods.
Tight convergence of the SGM algorithm can be challenging, because converging the gradient to a sufficient degree requires very tight convergence of the squared gradient. 
In our calculations, we used a rather loose convergence criterion of  $10^{-4}$.
The results are displayed in panels C and D of Fig.~\ref{fgr:DAscans} for the ALMO-SGM and FR-SGM methods, respectively.
The guess energy and $\mathrm{D}_\mathrm{CT}$ descriptors are shown as red circles, whereas the fully converged results are displayed as solid dots.
The ALMO guess energy increases when bringing the D and A monomers closer, in contrast to the expected asymptotic trend.
Clearly, however, the ALMO density is a suitable guess for further optimization with the SGM method.
This is demonstrated by the inset of Fig.~\ref{fgr:DAscans} panel C, where the results obtained from the guess density are close to the converged results and by the fact that the SGM optimization relaxed smoothly all data points, stabilizing some by several eVs.
Conversely, the constrained optimization (FR method, Fig.~\ref{fgr:DAscans} panel D) produces both energies and densities very close to the target stationary state.
The asymptotic trend is well reproduced in the dihedral scan for both OO-DFT methods, with the sole exception of the data point corresponding to a dihedral angle of 31.3\degree, which was found to be problematic also for other methods.
Evidently, the density obtained by the constrained FR optimization is already very close to the target state, meaning most of the density relaxation involves MOs that are not directly involved in the excitation, but are influenced by the charge separation in the CT state through electrostatic interaction.
The ALMO-SGM and FR-SGM methods lead to almost identical values for $I P_{\mathrm{D}}-E A_{\mathrm{A}}$ in the fit (rows 4 and 5 in Table \ref{tbl:ipea}), within 0.5 eV of the results obtained with the LRC-$\omega_\mathrm{GDD}$PBE reparametrization.
As a conclusion, we recommend the FR-SGM OO-DFT method for the calculation of ICT excitations in large D-A dimers, and the LRC-$\omega_\mathrm{GDD}$PBE reparametrization for TD-DFT calculations. \\

\begin{figure}[h!]
 \centering
 \includegraphics[width=0.95\textwidth]{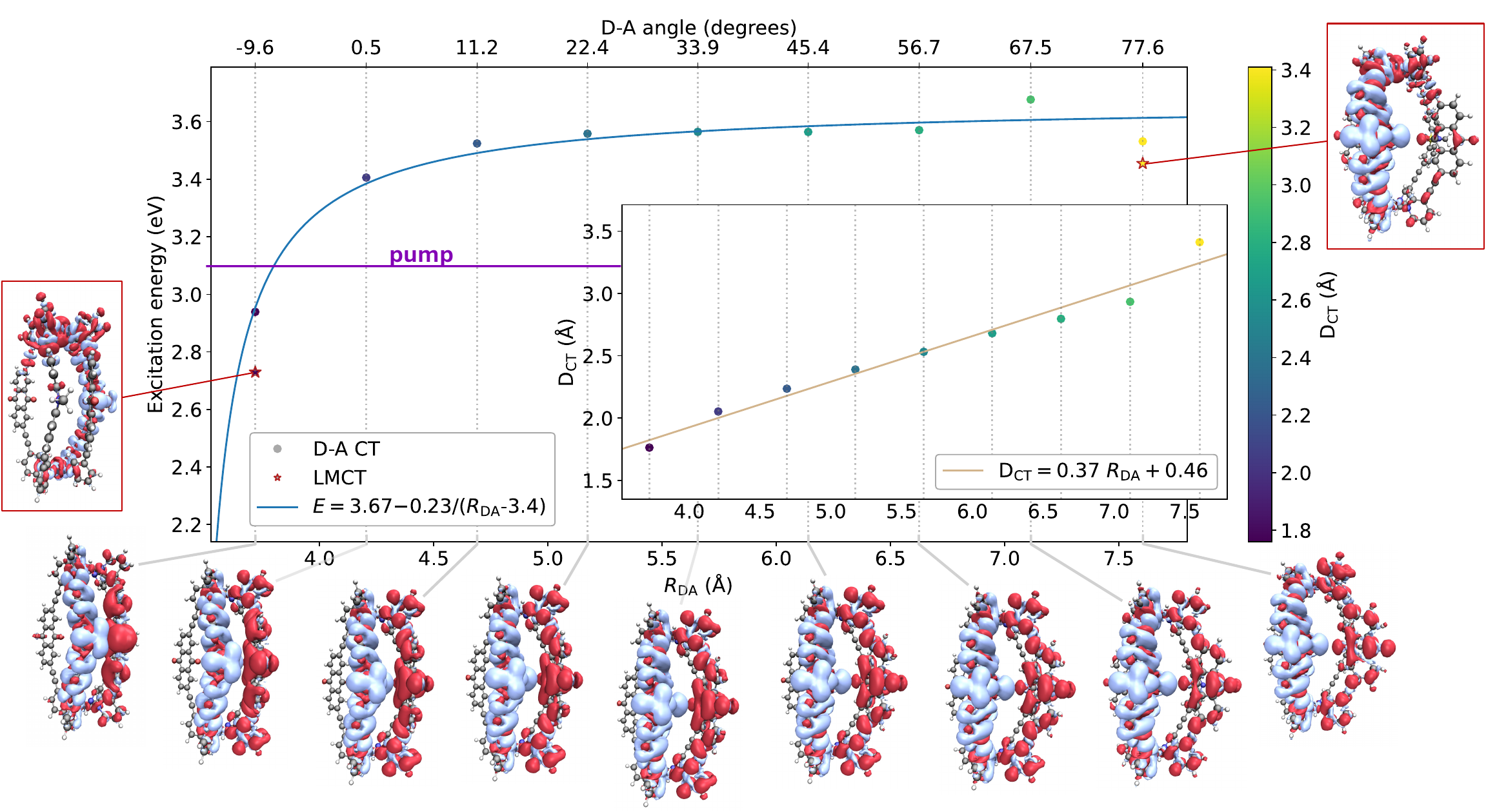}
 \caption{
 D-A CT excitations calculated with FR-SGM for the full cage structure as a function of the donor-acceptor distance. The density difference between the ground and ICT excited state is shown at the bottom of each panel as an isosurface plot (isovalue 0.0005 for the blue surface and -0.0005 for the red surface).
 The energies are fitted with a shifted asymptote (Eq.~\ref{eq:shifted}).
 The D-A CT excitations are marked with dots, and additionally calculated LMCT excitations are marked with a red-framed star.
 Inset: $\mathrm{D}_\mathrm{CT}$ values plotted against the donor-acceptor separation  distance $R_\mathrm{DA}$ and fitted with a linear expression.}
 \label{fgr:cage}
\end{figure}

\begin{figure}[h!]
\centering
  \includegraphics[width=0.85\textwidth]{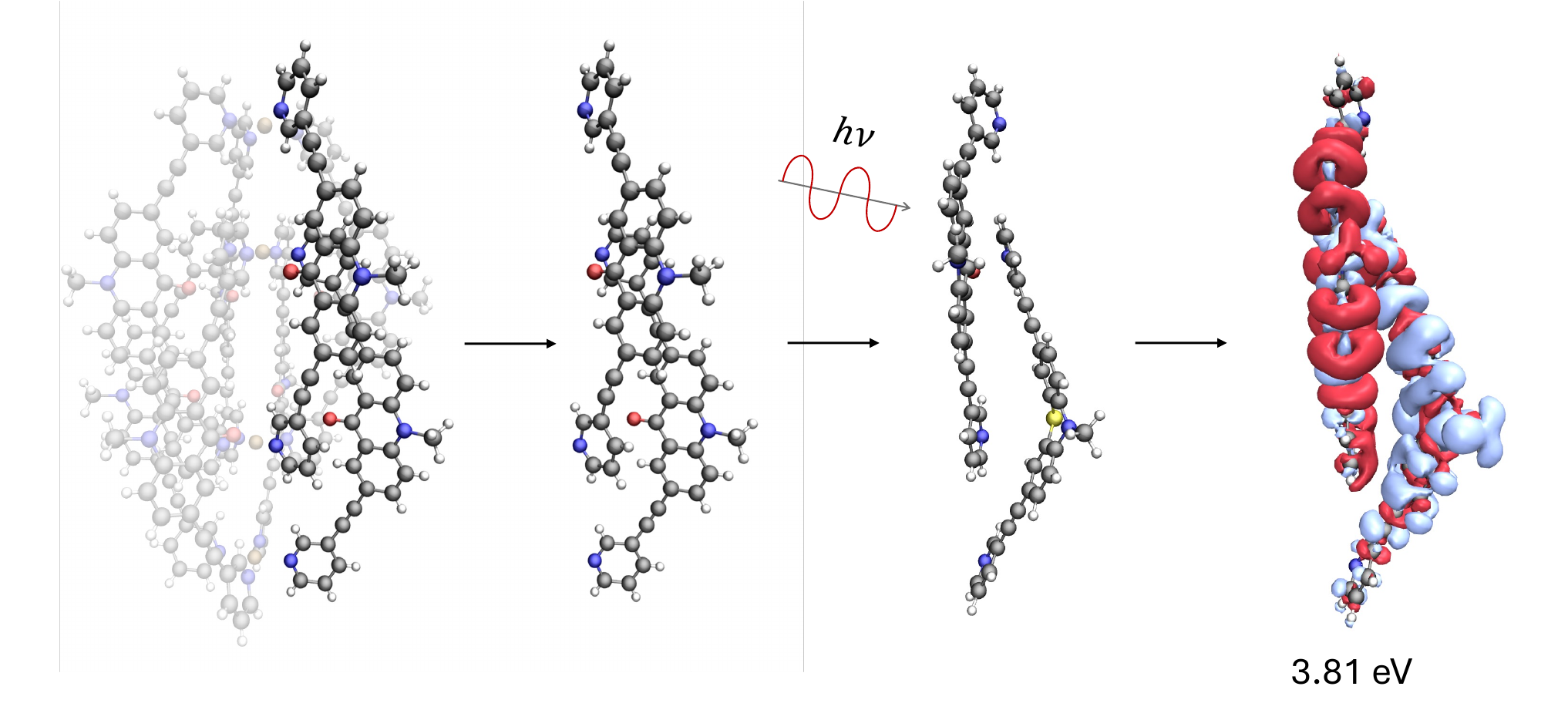}
  \caption{Starting from the crystal structure in Ref.~\citenum{frank2016light} an ANQ molecular dimer is isolated with monomers coming from two different cages of the interlocked structure.
  One of the ANQ ligands (acceptor) is then swapped for a PTZ (donor) ligand, preserving the local N-Pd coordination environment.
  A calculation of the PTZ-ANQ ICT excitation in this conformation is then performed with the ALMO-SGM method. The calculated excitation energy of 3.81 eV, is in the same energy range predicted for CT excitations in the single cage-derived donor-acceptor conformations (see Fig.~\ref{fgr:cage}).
  The predicted ICT excitation energy lies above the energy of the pump pulse used in Ref.~\citenum{frank2016light}.}
  \label{fgr:doublecage}
\end{figure}

We then applied the FR-SGM method to the calculation of D-A CT excitations in the full PTZ-ANQ Pd coordination cage.
The Pd coordination environment is conserved in the structures interpolated as described above and we computed the CT excitation energies for nine structures of the D-A dihedral scan.
The D-A dihedral in Fig.~\ref{fgr:cage} is obtained from planes fitted only through the dye cores (secondary x-axis).
By analogy, the $R_\mathrm{DA}$ separation (primary x-axis) is computed as the distance of the center of nuclear charge of the D and A dye cores.
The resulting dihedral spans a wider range compared to ligand-only study in Fig.~\ref{fgr:DAscans}, from -9.6\degree \, in the stacked conformer to 77.6\degree \,  in the orthogonal conformer.
The FR-SGM method converged reliably to the lowest-lying CT excitation involving the D and A dyes (circles).
For the full cage, we decided to fit the calculated ICT energies to a shifted expression
\begin{equation}\label{eq:shifted}
E=a-\frac{b}{R_\mathrm{DA}-c},
\end{equation}
where the $c$ parameter describes a shift of the x-axis intercept.
We specifically set $c=3.4 $ \r{A} which is still smaller than the D-A separation $R_\mathrm{DA}$ measured in the stacked conformer.
The IP-EA asymptote is lower by 0.5 eV compared to the D-A ligand-only calculations, so we expect that the CT excitation in the stacked conformation is accessible with a 400 nm pump pulse, as demonstrated in Ref.~\citenum{frank2016light}.
This means confinement of the D and A monomers in the supramolecular cage enables the formation of PTZ-ANQ CT excitations, but the cage environment in the stacked conformation slightly reduces the energy of the PTZ-ANQ CT excitation.
The $\mathrm{D}_\mathrm{CT}$ plot in the inset of Fig.~\ref{fgr:cage} shows the expected monotonous linear increase with the D-A separation.
No data points had to be excluded from the energy and $\mathrm{D}_\mathrm{CT}$ fits. \\
We also computed ``pure'' LMCT excitations (red star marker in Fig. \ref{fgr:cage}) for the stacked and orthogonal conformers.
Keeping in mind all uncertainties of our method, our calculations predict LMCT excitations in the same energy range as the D-A CT excitations, both in stacked and orthogonal conformations.
Hence, we cannot rule out excited-state relaxation to the LMCT excitations following the experimentally applied 400 nm pulse.\\
Finally, we estimate the likelihood of inter-cage CT within the interlocked cage dimer.
Starting from the crystallographic data reported in Ref.~\citenum{frank2016light}, we isolated an ANQ-ANQ dimer in a stacked conformation.
We then swapped one ANQ ligand for a PTZ ligand, aligning the N atoms to the Pd coordination sites in a $\pi$-stacked conformation.
The process we followed to generate this stacked conformation as an approximation to the interlocked cage structures is summarized in Fig.~\ref{fgr:doublecage}.
The ALMO-SGM calculation shows the presence of a charge-transfer excitation at 3.81 eV, confirming the presence of inter-cage excitations in the energy range of intra-cage CT excitations.
However, to confirm the assignment of the PTZ-ANQ excitation to an inter-cage CT would require the computationally prohibitively expensive calculation of the full interlocked cage dimer, since in our model calculation, this ICT state would not be accessible with a 400 nm pulse.
In conclusion, we confirm the presence of low-lying D-A CT and LMCT excitations in the supramolecular Pd coordination cage compounds.
We find inter-cage CT excitations in an energy range compatible with the cage conformations, but cannot exclude the formation of LMCT excitations in the excited-state deactivation process.

\subsection{Charge-transfer excitations in dye-TiO$_2$ complexes}
\label{sec:dye-TiO2}

We also tested the FR-SGM method on CT excitations in prototypical dye-semiconductor complexes of technological interest for their application in dye-sensitized solar cells (DSSC).
The injection of an excited electron from the organic dyes was previously simulated for the D102-TiO$_2$ and JK2-TiO$_2$ dye-semiconductor complexes by Gemeri \textit{et al.} using TD-DFT.\cite{gemeri2022electronic}
We used the molecular structures of the dye-TiO$_2$ complexes reported in Ref.~ \citenum{gemeri2022electronic}.
These systems are chemically different from the supramolecular cages since the charge acceptor is a semiconductor modeled as a cluster.
The Orca software package\cite{neese2022software} was used to perform a preliminary simplified TDA (sTDA) calculation of the low-lying excitations in the dye-semiconductor complexes, using the $\omega$B97X-D XCF and def2-TZVP basis.
Low-lying CT excitations were then optimized using the FR-SGM method and the same XCF and basis set.
The sTDA and FR-SGM results are compared in Table \ref{tbl:dyes-tio2}.

\begin{table}[h!]
\small
  \caption{Excitation energies for low-lying charge-transfer excitations in D102-TiO$_2$ and JK2-TiO$_2$ dye-semiconductor complexes, computed with the sTDA method in Orca and the OO-DFT method in Q-Chem.
  All energies are given in eV and $\mathrm{D}_\mathrm{CT}$ is given in \AA.}
  \label{tbl:dyes-tio2}
  \begin{tabular*}{0.78\textwidth}{@{\extracolsep{\fill}}llllll}
    \hline
Dye & Dominant excitation & sTDA & FR-SGM & $\mathrm{D}_\mathrm{CT}$ & class \\
    \hline
D102 & 881-896 & 3.48 & 5.45 & 3.21 & dye-TiO$_2$ \\
JK2  & 915-916 & 2.84 & 2.56 & 6.01 & dye-spacer  \\
JK2  & 914-916 & 4.03 & 3.17 & 5.90 & dye-spacer  \\
JK2  & 915-917 & 4.61 & 3.50 & 7.45 & dye-TiO$_2$ \\
JK2  & 915-918 & 4.69 & 3.57 & 7.41 & dye-TiO$_2$ \\
JK2  & 915-919 & 4.65 & 3.23 & 6.59 & dye-TiO$_2$ \\
JK2  & 915-920 & 4.75 & 3.76 & 7.48 & dye-TiO$_2$ \\
JK2  & 915-921 & 4.80 & 3.82 & 7.52 & dye-TiO$_2$ \\
JK2  & 915-922 & 4.95 & 3.72 & 7.33 & dye-TiO$_2$ \\
JK2  & 915-923 & 4.99 & 3.91 & 7.61 & dye-TiO$_2$ \\
    \hline
  \end{tabular*}
\end{table}

\begin{figure}[h!]
\centering
  \includegraphics[width=0.9\textwidth]{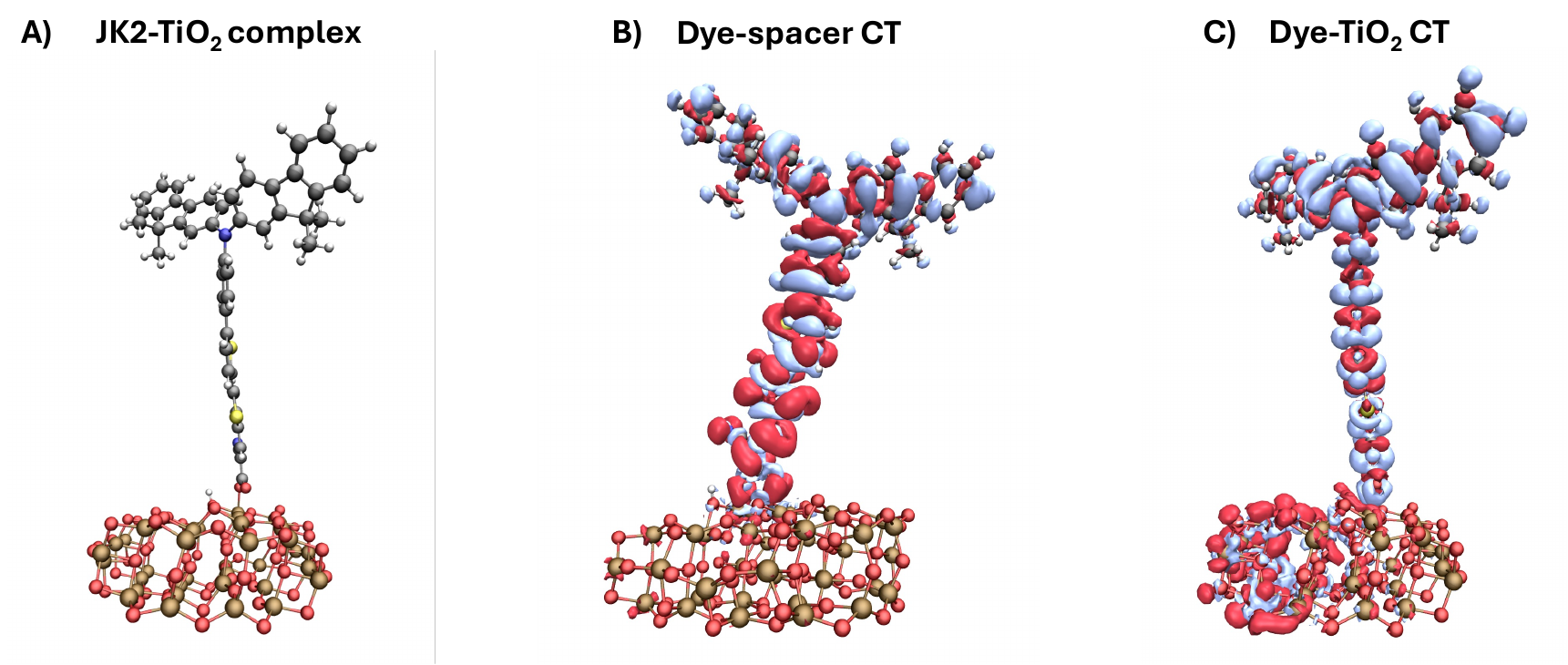}
  \caption{Structure of the JK2-TiO$_2$ dye-semiconductor complex (panel A). Isosurface plot of the density differences for dye-spacer (panel B), and dye-TiO$_2$ (panel C) CT excitations of the JK2-TiO$_2$ complex computed with the FR-SGM method. The isosurface plot is obtained by using an isovalue of 0.0005 for the blue surface and -0.0005 for the red surface.}
  \label{fgr:jk2-Tio2-new}
\end{figure}

One direct dye-TiO$_2$ excitation was calculated with the FR-SGM method for the D102-TiO$_2$ complex, which appears strongly blue-shifted compared to the sTDA result.
On the contrary, all excitations computed with FR-SGM in the JK2-TiO$_2$ complex are slightly red-shifted in comparison to the sTDA result.
The JK2 ligand is characterized by a long aromatic ``spacer'' connecting the dye to the semiconductor.
By inspecting the density differences, we classified all CT excitations computed in Table~\ref{tbl:dyes-tio2} as ``dye-spacer'' or ``dye-TiO$_2$''.
The JK2-semiconductor complex shows two groups of CT states: dye-spacer excitations reside in the visible spectrum, and dye-TiO$_2$ excitations, resulting in the direct injection of a dye electron in the semiconductor band, populate the near-UV spectral region.
Fig. ~\ref{fgr:jk2-Tio2-new} shows prototypical density difference plots for both groups of CT states.

\section{Conclusions}

In this work, we identified ALMO calculations for ionized fragments and a constrained optimization (FR) procedure similar to the method reported by Schmerwitz \textit{et al.}\cite{schmerwitz2025freeze} to yield accurate guesses density for the calculation of ICT excitations with OO-DFT.
We evaluated the performance of these guesses on ICT excitations in the tetrafluoroethylene-ethylene dimer system from the $R_\mathrm{DA}$-dataset\cite{D4CP01866D}, with IMOM and SGM OO-DFT methods.
The FR-guess density provides an excellent guess for OO-DFT calculations of ICT excitations when coupled to the SGM algorithm (FR-SGM method) which is also true for the ALMO guess and the corresponding ALMO-SGM method which we recommended whenever the separation into discrete fragments is straightforward and the donor and acceptor fragments are not connected by a covalent bound.\\
We further compared these two OO-DFT methods to system-specific reparametrizations of the LRC-$\omega$PBE XCF in the TDA calculation of ICT in molecular dimer systems for several conformers of a donor-acceptor ligand system which was cut out from a supramolecular cage structure.
Both the ALMO and FR guess densities were shown to be adequate, with the FR guess being closer to the target energy.
The global density-dependent tuning\cite{modrzejewski2013density, mandal2025simplified} of the range-separation parameter $\omega_\mathrm{GDD}$ also performed very well for these systems.
In contrast, the parametrization based on the excited-state density descriptor $\mathrm{D}_\mathrm{CT}$\cite{yan2025adaptable} results in irregular $\omega$-values and ICT energies.
We hence recommend the size-extensive LRC-$\omega_\mathrm{GDD}$PBE method for the calculation of ICT excitations with TDA or suggest the $\mathrm{D}_\mathrm{CT}$-based tuning procedure to only take the asymptotic value at large distances into account. \\
The promising FR-SGM method was used for the calculation of CT excitations in three large supramolecular systems.
Our calculations of PTZ-ANQ excitations in the full coordination cage investigated experimentally by Frank \textit{et al.}\cite{frank2016light} predicted excitation energies in the visible range for the stacked conformation, as well as LMCT excitations.
To confirm the true character of the experimentally observed excitations, further calculations of the intercalated double-cage dimer are required but are computationally very demanding even for the scalable methods identified in this work.
Finally, CT excitations in two covalently bonded dye-semiconductor complexes, D102-TiO$_2$ and JK2-TiO$_2$,\cite{gemeri2022electronic} were computed with the sTDA and FR-SGM methods.
FR-SGM reliably converged on the CT excitations also for these systems. \\
OO-DFT methods hold great potential for the simulation of photoinduced phenomena in supramolecular systems, due to their low computational scaling and balanced description of CT excitations, whereas TDA requires system-specific XCF parametrizations.
The implementation of a robust guess refinement method establishes a recipe for the reliable convergence of OO-DFT calculations, enabling further investigation of semi-empirical XCFs for the balanced description of ground and excited states.
Further development is desirable for the convergence acceleration of the SGM algorithm. 


\section*{Conflicts of Interest}
M.H.G. is a part-owner of Q-Chem Inc., whose software was used for all developments and calculations reported here.

\section{Acknowledgments}

N.B. thanks Gabriel Jonathan Flür for performing initial calculations on the dye-TiO$_2$ systems.
Financial support by the DFG under Germany's Excellence Strategy - EXC 2089/1-390776260 (e-conversion) is gratefully acknowledged.
A research stay of N.B. at UC Berkeley was funded by the Deutsche Forschungsgemeinschaft (DFG) through the Research Training Group “Confinement Controlled Chemistry” (GRK 2376). Work at Berkeley was funded by the DOE Office of Science, Basic Energy Science (BES) Program, Chemical Sciences, Geosciences and Biosciences Division under Contract no. DE-AC02-05CH11231, through the Atomic, Molecular, and Optical Sciences program.

\providecommand{\latin}[1]{#1}
\makeatletter
\providecommand{\doi}
  {\begingroup\let\do\@makeother\dospecials
  \catcode`\{=1 \catcode`\}=2 \doi@aux}
\providecommand{\doi@aux}[1]{\endgroup\texttt{#1}}
\makeatother
\providecommand*\mcitethebibliography{\thebibliography}
\csname @ifundefined\endcsname{endmcitethebibliography}
  {\let\endmcitethebibliography\endthebibliography}{}


\begin{figure}[H]
\centering
  \includegraphics[width=0.75\textwidth]{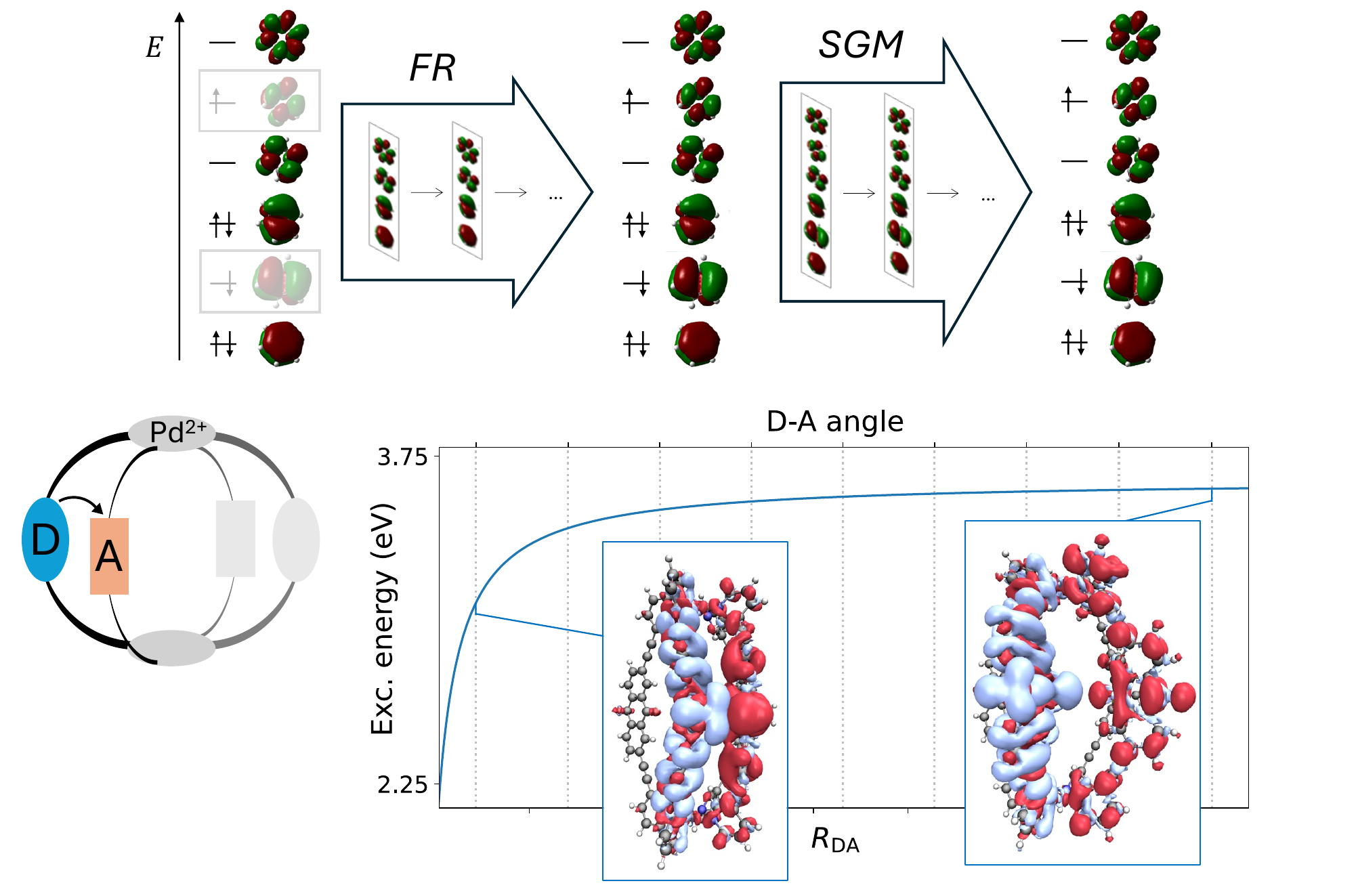}
  \caption{TOC graphic}
  \label{fgr:toc}
\end{figure}

\end{document}